# Deposited low temperature silicon GHz modulator

Yoon Ho Daniel Lee[1], Michael O. Thompson[2], Michal Lipson[1,3,*]

[1]Cornell Nanophotonics Group, School of Electrical and Computer Engineering,
Cornell University, Ithaca, New York 14853, USA

[2]Department of Materials Science and Engineering,
Cornell University, Ithaca, New York 14853, USA

[3]Kavli Institute at Cornell, Cornell University, Ithaca, New York 14853, USA
*ml292@cornell.edu

The majority of silicon photonics is built on silicon-on-insulator (SOI) wafers while the majority of electronics, including CPUs and memory, are built on bulk silicon wafers, limiting broader acceptance of silicon photonics. This discrepancy is a result of silicon photonics' requirement for a single-crystalline silicon (c-Si) layer and a thick undercladding for optical guiding that bulk silicon wafers do not provide. While the undercladding problem can be partially addressed by substrate removal techniques[1, 2], the complexity of co-integrating photonics with state-of-the-art transistors[3] and real estate competition between electronics and photonics remain problematic. We show here a platform for deposited GHz silicon photonics based on polycrystalline silicon with high optical quality suitable for high performance electro-optic devices.  We demonstrate 3 Gbps polysilicon electro-optic modulator fabricated on a deposited polysilicon layer fully compatible with CMOS backend integration. These results open up an array of possibilities for silicon photonics including photonics on DRAM and flexible substrates.

Previous attempts at integrating silicon photonics with bulk CMOS include localized substrate removal[1] and bonding techniques[4], but process compatibility, scalability and manufacturability have hindered these approaches from mainstream adoption. Localized substrate removal involves using $XeF_2$ gas to isotropically undercut the bulk silicon substrate underneath the waveguide to prevent optical leakage into the substrate. However, it only allows fabrication of waveguides from the transistor gate polysilicon layer, which cannot be independently optimized for optical quality. A germanium electro-absorption modulator has also been demonstrated[5], but it requires a crystalline silicon layer as well as thermal processing at 550 ˚C. Other approaches involve transfer or bonding of thin films, individual devices, or complete dies. Bonding of unpatterned crystalline silicon[4] allows single crystalline material on CMOS BEOL, but it does not reduce the high thermal budget required for dopant activation to fabricate active devices. For transfer of completely fabricated devices[6], yield and alignment tolerance remain challenging. Flip-chip bonding of a complete SOI photonic die[7,8] onto an electronics die has also been used due to the maturity of flip-chip bonding technology. However, flip-chip bonding suffers from additional electrical parasitics, limited architectural freedom, and high cost.

There are several low loss optical materials that can be deposited and in principle be integrated with CMOS, but their high band gap and low mobility precludes these materials from enabling active devices such as modulator and switches. Examples of such materials include polymers[9], Silicon Nitride (SiN)[10,11] and hydrogenated amorphous silicon (a-Si:H)[12]. SiN is a dielectric with bandgap of 5 eV making it electrically inactive, and a-Si:H with its inherently high defect density and low mobility requires high voltage incompatible with latest CMOS transistors and is unable to operate at gigahertz speed. Polymer has low upper thermal processing limit, and high poling voltage (> 10 V) is needed for activation of polymer, posing compatibility and reliability issues for system integration. Furthermore, active polymer devices require c-Si waveguide for its operation, which makes backend integration difficult.

Polysilicon is a deposited material that could enable high performance active devices, but it traditionally exhibited much lower performance than its crystalline counterpart. Polysilicon is a collection of grains of c-Si separated by grain boundaries consisting of a few atomic layers of amorphous material. Grain boundaries not only act as small perturbations causing photon and electron scattering, but also create states within the silicon bandgap that cause excess optical loss. Since the groundbreaking work on polysilicon photonics[13], much progress has been made, including active devices. Various works using high temperature annealed polysilicon have demonstrated low loss on the order of 10 dB / cm[14], and electro-optic modulation[15, 16] has also been shown. However, these works are not compatible with CMOS backend integration due to their high thermal budget fundamental to furnace annealed polysilicon, constraining them to frontend integration in CMOS and DRAM.

We demonstrate high performance deposited silicon photonics on a thin film of low temperature polysilicon by tailoring the dimensions of the grain boundaries to be similar to the dimensions of the cross-sections of nanophotonic devices. By ensuring that the number of grain boundaries across the cross-section of the waveguide is small, the electrical properties of the device are expected to be comparable to its single crystalline counterpart, and the optical properties to be sufficient for high quality factor resonators. The tailoring of the grain sizes is done by using Excimer Laser Anneal (ELA), which involves irradiation of a deposited thin film of a-Si with a short UV laser pulse, which is absorbed completely within the first tens of nm due to high absorption coefficient of a-Si at UV wavelengths. Due to the very short duration and confinement of heat close to the surface, underlying device structures receive a minimal increase in the total thermal budget, while the a-Si film heats up enough to melt and re-solidify, crystallizing into polysilicon. In fact, the thermal impact is so small that ELA has been successfully performed even on a plastic substrate with 120 ˚C thermal processing limit[17]. This local heating enables ELA to be performed in the CMOS backend without affecting the electronics underneath, decoupling the CMOS frontend from photonics (Fig. 1). Furthermore, excimer laser annealing is an industry proven technology in the flat panel industry with throughput equivalent to over five hundred 300 mm wafers per hour, well exceeding that of state of the art CMOS lithography tools[18]. Once ELA is performed on the wafer, the rest of the fabrication is performed using standard CMOS fabrication process, therefore requiring only one extra process module in a CMOS process.

We fabricate low temperature polysilicon ring resonators utilizing one step crystallization as described in the Methods and show high performance resonators suitable for up to 15 GHz modulation bandwidth with high extinction ratio. The thermal budget of the whole process flow is compatible with CMOS backend integration, with single dominant thermal budget of 1 hour at 450 ˚C for dehydrogenation. This thermal budget can in principle be significantly lowered by using multiple-step recrystallization[17] or evaporated material. The waveguide dimensions are 700 nm wide by 110 nm high with slab thickness of 40 nm for single mode transverse electric (TE) polarization operation. The waveguide has a high effective confinement factor of 0.78, which allows bending radius as tight as 5 μm yielding measured free spectral range of 26 nm. The low temperature ring resonators with radius of 20 μm have loaded Q-factor of 12,000, and 10 dB extinction ratio as shown in Fig. 2.(b). Note that the loss of the polysilicon (estimated to be on the order of 20 dB / cm) limits only the extinction ratio (currently 10dB) and not the insertion loss, due to the small length of the device.

We fabricate the modulators as described in the methods and show that ELA polysilicon has good dopant activation characteristics and c-Si-like behavior. The completed device is shown in Fig. 2.(a). We measured the IV characteristics of polysilicon PN diode ring modulators with 20 μm radius and observed total series resistance of 25 Ω and low reverse leakage current of -62 nA at -5 V. The diode IV curve seen in Fig. 2.(c) clearly shows exponential behavior in the low current regime below 0.8 V with diode ideality factor of 1.35 ± 0.1, followed by high injection and series resistance limited behavior.

Ideality factor of 1.35 along with low normalized leakage current of -490 pA / μm demonstrate that ELA polysilicon has great dopant activation characteristics and crystalline silicon-like behavior that make this diode well suited for sensitive forward bias modulation. In addition, combination of low leakage current and low series resistance would make ELA polysilicon an excellent platform for high speed reverse bias modulator.

We observed an open eye diagram up to 3 Gbps using pseudo random bit sequence (PRBS) $2^7$-1 pattern with pre-emphasis. In carrier injection mode, we measured electro-optic (EO) 10 % - 90 % rise time of ~500 ps and 90 % - 10 % fall time of ~400 ps using 2 $V_{p-p}$ square wave input signal with DC bias of 1.8V, as shown in Fig. 3(a). The rise and fall time values limit intrinsic EO bandwidth of the modulator to below 1GHz, as expected with silicon carrier injection modulators. In order to increase the bandwidth of the resonator, we applied 2 $V_{p-p}$ PRBS 2^7-1 pattern with ± 1.5 V pre-emphasis and 1.2 V DC bias to the modulator (similar to what is done in the case of single crystalline silicon modulator[19]) and measured open eye diagram at 3 Gbps at an operating wavelength of 1598.9 nm, as shown in Fig. 3(b). The eye opening was limited due to lithographic misalignment between the heavily doped regions and the waveguides.

In conclusion we have demonstrated a low temperature, deposited silicon ring modulator operating at 3 Gbps using excimer laser annealing. Low thermal budget and strict adherence to standard CMOS fabrication makes this novel platform compatible with CMOS backend integration. In addition, this method frees silicon photonics from its dependence on SOI and allows true monolithic integration with bulk CMOS electronics, DRAM, and even flexible substrates. Most importantly it decouples CMOS frontend from photonics, which lowers the barrier of introducing silicon photonics into traditional CMOS foundries, leading to rapid adoption of photonics.

**Methods**

We start fabrication on a silicon wafer with 4 µm of thermal oxide. Deposition of 150 nm of undoped PECVD a-Si under 400 ˚C is performed by a commercial deposition service. A series of electron beam lithography and ion implantation is used to form N, N++, P, and P++ regions using Phosphorous and Boron, respectively. The wafer is then furnace annealed at 450 ˚C for 1 hour in argon ambient to dehydrogenate the PECVD a-Si film, followed by excimer laser annealing using a laser pulse at wavelength of 308 nm with temporal FWHM of 35 ns generated by a XeCl excimer laser. This step crystallizes the initial layer of deposited a-Si into polysilicon and makes the dopants electrically active. Following ELA, excess surface roughness created by ELA is polished using Chemical Mechanical Polish (CMP). Waveguide and slab are defined by electron beam lithography and etched using reactive ion etching. The wafer is then clad with 1 µm of PECVD $SiO_2$ at 400 ˚C, followed by electrical contact and pad formation.

**Acknowledgement**

This work was supported in part by DARPA award # W911NF-11-1-0435 supervised by Dr. Jagdeep Shah, Intel Corporation, and NSF through CIAN ERC under Grant EEC-0812072. This work was also supported in part by NSF and Semiconductor Research Corporation under Grant ECCS-0903406 SRC Task 2001, and NSF Grant #1143893. This work was performed in part at the Cornell NanoScale Facility, a member of the National Nanotechnology Infrastructure Network, which is supported by the National Science Foundation (Grant ECS-0335765).

**Author Contribution**

Y.H.D.L. simulated, designed, and fabricated the sample and conducted the experiments. Y.H.D.L. and M. O. T. discussed the Excimer laser annealing process. Y.H.D.L., M.O.T. and M.L. discussed the results and implications.

**Competing financial interests**

The authors declare no competing financial interests.

**Figure Legends**

Fig. 1. Backend deposited silicon photonics. Rendered image of polysilicon modulator integrated on CMOS BEOL. For clarity, we show only a part of the metal contacts. One can see that the grain boundaries and the dimensions of the cross-section of the device are comparable.

Fig. 2. Characterization of low temperature polysilicon devices. (a) Optical micrograph of the fabricated device (b) Transmission spectrum of polysilicon ring resonator with $Q_{loaded}$ ~12,000 (c) IV curve of the fabricated polysilicon ring modulator device.

Fig. 3. Electro-optic modulation using polysilicon modulator. (a) Modulator output with square wave input signal. (b) Optical eye diagram of polysilicon ring modulator at 1598.9 nm (PRBS $2^7$-1 pattern with pre-emphasis at 3 Gbps ).

**Figures**

**Figure 1**

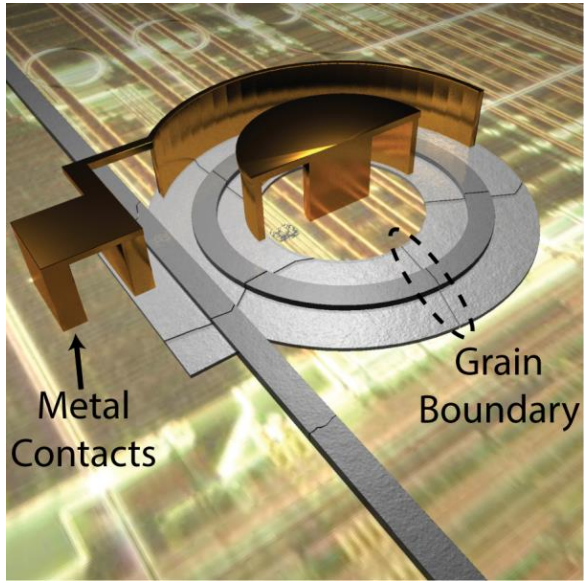

**Figure 2**

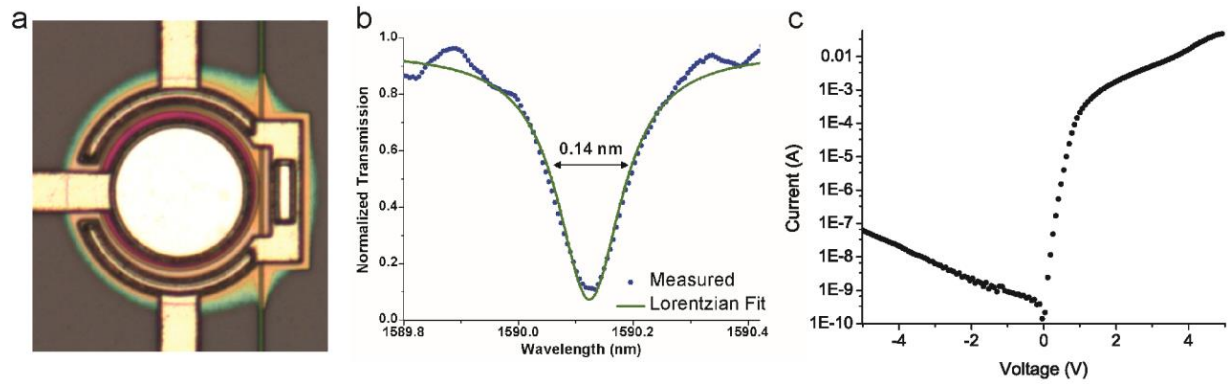

**Figure 3**

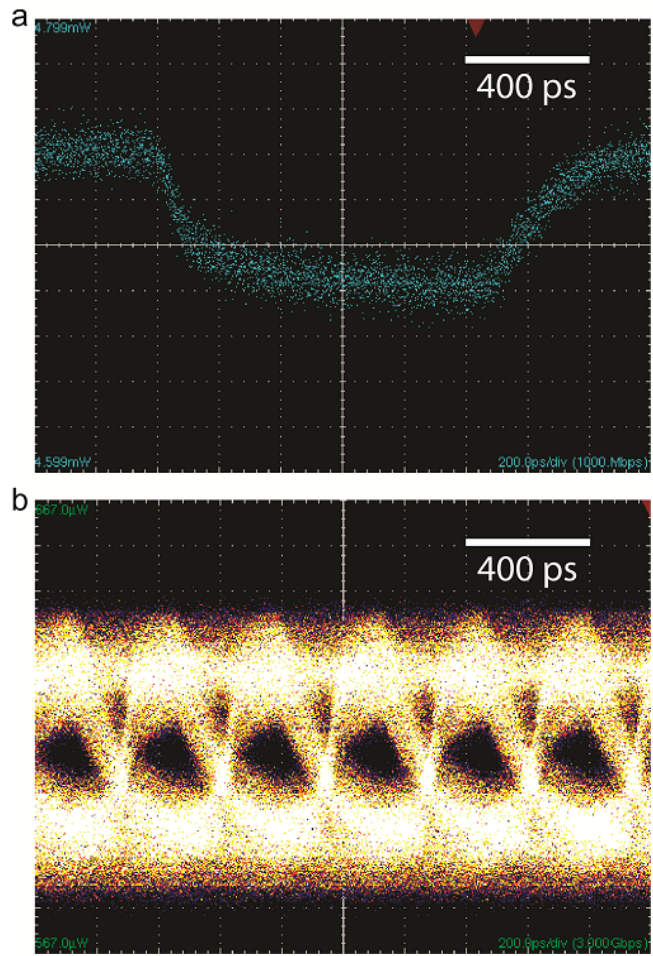